\documentclass[a4paper,11pt]{article}

\usepackage{aas_macros}
\usepackage{pos}

\title{ ``Spider'' Millisecond Pulsar Binaries as Potential TeV Emitters}

\author*[a,b,c]{Zorawar Wadiasingh}

\author[c]{Christian J. T. van der Merwe }

\author[c]{Christo Venter}

\author[d]{Alice K. Harding}

\author[e]{Matthew G. Baring}

\affiliation[a]{Astrophysics Science Division, NASA Goddard Space Flight Center, Greenbelt, MD 20771, USA}
\affiliation[b]{Universities Space Research Association (USRA), Columbia, MD 21046, USA}
\affiliation[c]{Centre for Space Research, North-West University, Potchefstroom Campus, Private Bag X6001, Potchefstroom 2520, South Africa}
\affiliation[d]{Theoretical Division, Los Alamos National Laboratory, Los Alamos, NM 58545, USA}
\affiliation[e]{Department of Physics and Astronomy - MS 108, Rice University, 6100 Main St., Houston, TX 77251-1892, USA}

\emailAdd{zwadiasingh@gmail.com}

\abstract{Pulsar winds interacting with sources of external pressure are well-established as efficient and prolific TeV accelerators in our Galaxy. Yet, enabled by observations from Fermi-LAT, a growing class of non-accreting pulsars in binaries has emerged and these are likely to become apparent as TeV emitters in the CTA era. This class consists of the black widows and redbacks,  binaries in which a millisecond pulsar interacts with its low-mass companion. In such systems, an intrabinary shock can form as a site of particle acceleration and associated nonthermal emission. We motivate why these sources are particularly interesting for understanding pulsar winds. We also describe our new multizone code which models the X-ray and gamma-ray synchrotron and inverse Compton spectral components for
select spider binaries, including diffusion, convection, and radiative
energy losses in an axially symmetric, steady-state approach. This new
multizone code simultaneously yields energy-dependent light curves and
orbital-phase-resolved spectra. It also better constrains
the multiplicity of electron/positron pairs that have been accelerated
up to TeV energies and are necessary to power orbitally-modulated
synchrotron emission components between the X-rays and MeV/GeV bands
potentially observed in some systems. This affords a more robust prediction
of the expected high-energy and VHE gamma-ray flux. Nearby MSPs with hot or flaring companions may be promising targets
for CTA, and it is possible that spider binaries could contribute to the observed AMS-02 energetic positron excess.

}

\FullConference{%
  *** 37th International Cosmic Ray Conference (ICRC2021), ***\\
  *** 12-23 July 2021 ***\\
  *** Berlin, Germany - Online ***
}

\begin{document}
\maketitle

\section{Background}

Pulsars and interaction of pulsar winds with sources of external pressure are well-established as efficient and prolific TeV accelerators in our Galaxy, as revealed by H.E.S.S., MAGIC, VERITAS and HAWC. Among these are classical gamma-ray binaries, such as PSR B1259--63, wherein an interaction of a young pulsar's wind with the companion's wind or excretion disk results in particle acceleration and orbitally-modulated high-energy emission at an intrabinary shock. Yet, enabled by observations from Fermi-LAT, a growing class of non-accreting pulsars in binaries has emerged and these are likely to be TeV emitters that are detectable using CTA. These are the ``spider'' binaries where a millisecond pulsar (MSP) \cite{2021arXiv210105751H} in a highly compact circular binary (orbital periods $\lesssim 1-2$ days) interacts with a low-mass companion. Two families of these binaries exist, distinguished by the nature of the companion: the black widows (with companion mass $m_c \lesssim 0.1 M_\odot$) and redbacks ($m_c \gtrsim 0.1 M_\odot$) \cite{2011AIPC.1357..127R}. The number of these systems has grown from the singular Black Widow pulsar PSR B1927+20, to over 20 in the past decade (including ``transitional" millisecond pulsars which switch between rotational and accretion powered states; here we are concerned with systems in the rotational-powered ``pulsar state"). For recent lists of candidates, see \cite{2015ApJ...807..130V,2019ApJ...872...42S,2020arXiv200403128T,2021JCAP...02..030L,2020arXiv201009060P} and the \href{https://confluence.slac.stanford.edu/display/GLAMCOG/Public+List+of+LAT-Detected+Gamma-Ray+Pulsars}{public {\it Fermi} LAT pulsar list}.

An unexpected (but now well-established) feature of many redbacks is that the intrabinary shock enshrouds the MSP, i.e. the companion's pressure dominates (in contrast to the canonical picture in some black widows) \cite{2015hasa.confE..29W,2016arXiv160603518R,2017ApJ...839...80W}. This paradigm shift in understanding these systems has occurred in the last $\sim 5$ years, enabled by detailed follow-up X-ray characterization and phasing of orbitally-modulated nonthermal X-ray emission.  These nonthermal X-rays originate from particles  accelerated at the intrabinary shock that principally cool via the synchrotron radiation channel, with Doppler-boosting along the shock from anisotropic particle distributions in the observer frame. Double-peaks in X-rays pulses arise as the Doppler-boosted flow sweeps past an observer's line-of-sight, with the two peaks corresponding to the wings of the shock.  Such a shock orientation is also revealed by orbital-phase and frequency-dependent radio eclipses of the MSP, where in RBs the eclipse fraction can be $> 50\%$ \citep[e.g.,][]{2009Sci...324.1411A} of the orbital phase, while the MSP remains largely uneclipsed at inferior conjunction of the pulsar.   This unexpected shock orientation and geometry, which implies that the companion overpowers the MSP's wind, possibly originates due to a strong companion magnetosphere or irradiation-stabilized radiatively inefficient accretion flow \citep[e.g.,][]{2018ApJ...869..120W}.

 As a consequence of the pressure balance and shock orientation, a large fraction of the pulsar's total wind output is captured by the shock, enabling a high radiative efficiency for a given spin-down power (a few percent in the classical soft X-ray band, cf.~\cite{2015arXiv150207208R}). This is important for subsequently heating/irradiating of the companion and also boosting the expected inverse Compton TeV flux.

\subsection{Why are ``spider'' binaries interesting and useful from a theoretical perspective?}

Several facets enable ``spider'' binaries to be potentially useful astrophysical laboratories of pulsars and pulsar winds. The intrabinary shock is essentially a pulsar wind termination shock on tiny scales of orbital separation $a \sim 10^{11}$ cm. The orbital modulation in X-rays through TeV offers a compelling diagnostic of oblique pulsar wind shock structure and the accelerated anisotropic particle distributions.
\newpage
 Moreover:
\begin{itemize}
\item Unlike pulsar $\gamma$-ray binaries, which are all in eccentric orbits (often with long intervals between periastrons), the short nearly circular orbits of spiders afford the study of the pulsar termination shock likely without the confounding factor of an orbitally-varying shock structure or geometry. 
\item The pulsar often has a well-timed ephemerides, and the optical companion mass function is typically characterizable. The optical modulation of the companion may also be modeled, thereby fitting for the system inclination. Therefore, all relevant orbital parameters and component masses are well-constrained or measured.
\item The MSP may be energetic, exceeding $10^{34}-10^{35}$ erg s$^{-1}$ spin-down power. Moreover, the pulsar's maximum potential drop, system Hillas criterion, and synchrotron radiation reaction all permit electrons with Lorentz factors $10^7-10^8$, i.e. multi-TeV energies. 
\item The proximity of the irradiated stellar companion, particularly in redbacks, implies that large target photon energy densities $u_\gamma \gtrsim 1$ erg cm$^{-3}$ are realized. These photons are an abundant target for inverse Compton upscattering to TeV energies. 
\item X-ray observations of the nonthermal synchrotron component ``anchor'' the particle distribution power-law and injection parameters for the spectral energy distribution. Combined with the known optical characterization of the companion, readily predictable TeV fluxes from inverse Compton emission are an automatic consequence.
\end{itemize}

 However, the X-ray emission alone does not pin down the maximum energy of the accelerated particles (given the unknown magnetic field at the shock), nor the total fraction of the pulsar spin-down power that goes into shock acceleration and emission. If the particle acceleration efficiency is high, spiders may also significantly contribute \citep{2015ApJ...807..130V,2021JCAP...02..030L} to the anomalous rise in the Galactic energetic positron fraction observed in low-Earth orbit by AMS-02, {\textit{PAMELA}}, and {\it{Fermi}}-LAT. Upper limits or detection by TeV telescopes of $\gamma$-ray emission from inverse Compton scattering of the relativistic shock-accelerated particles on the UV radiation of the hot irradiated companion stars could constrain the properties of the pulsar wind, shock magnetic field and of pulsar electron-positron pair multiplicity.

\section{Models}

Geometrical models \cite[e.g.,][]{2015hasa.confE..29W,2016arXiv160603518R,2017ApJ...839...80W}) can generally explain the morphology of X-ray light curves well via Doppler-boosting by bulk flow along the shock. Such geometrical models produce emission patterns for synchrotron emission as a function of binary inclination $i$ and other parameters, but make no predictions about spectral energy distribution (such as its normalization, or orbital phase dependence).

We have developed a new multi-zone code {\tt{umbrela}}, detailed in \cite{2020ApJ...904...91V}, which solves for the steady-state particle distributions (via an approximate form of the Boltzmann transport or convection-diffusion equation) in spatial zones along the shock with variable bulk motion, and considers synchrotron and inverse Compton channels for radiative losses. Particles are injected with a power law, $Q E^{-\Gamma}$, with $\Gamma$ varying to match X-ray spectra, with minimum, $E_{\rm min}$, and maximum, $E_{\rm max}$, cutoff energies, and normalized to a fraction of the pulsar power, similar to \citep{2015ApJ...807..130V}.  {\tt{umbrela}} is the first code which simultaneously and self-consistently models the X-ray and $\gamma$-ray light curves and spectra for spiders. The code takes as input known binary and pulsar parameters (including orbital inclination, orbital period, spin-down power, mass ratio), distance constraints, optical characterization of the system (radius, effective temperature), and particle spectral index (which may be readily chosen to be consistent with X-ray spectral index). Free parameters are those associated with the particle injection and shock, such as bulk motion, magnetic field at the shock, shock radius and geometry, acceleration efficiency, and pulsar pair multiplicity in terms of the Goldreich-Julian rate. 

 More specifically, we assume a hemispherical intrabinary shock that is subdivided into multiple spatial zones. For simplicity, the shock is assumed to possess azimuthal symmetry. In each zone, a parameterized fraction of the pulsar wind's power is injected as a power-law accelerated leptonic particle population. The contribution of particles advected from upstream zones along the shock are taken into account with additional injection terms. Energy losses are treated in each individual spatial zone for different effects (radiative losses, convection/adiabatic losses and diffusion), and a particular zone's proximity to the photon bath from the companion is treated appropriately. Note that synchrotron self-Compton radiation is generally negligible because the magnetic field is sufficiently low, and synchrotron and (external) inverse Compton losses dominate radiative losses.  We find that adiabatic losses generally dominate radiative ones, so the injected particle spectral shape does not change significantly across most zones. At the conclusion of the transport calculation, steady-state particle distributions in each zone are obtained, as well as the accelerated electron/positron population which escapes the system.  Emission from synchrotron and inverse Compton channels and Doppler boosting from the shock bulk flow are then computed for each zone, and summed to arrive at a total flux for a given orbital phase and observer viewing angle.

The upshot of such a formalism is that the code can predict spectra and phase-resolved orbital light curves simultaneously in an efficient manner, along with the cosmic ray pair flux supplied to the interstellar medium. The simplified transport treatment enables efficient sampling of the parameter space, setting the stage for future Markov Chain Monte Carlo exploration for a larger population of spider binaries.

\section{Recent Results}

In \cite{2020ApJ...904...91V}, we presented case studies of select spiders (those accessible to H.E.S.S.-II, but by no means a comprehensive study of all sources) with extant nonthermal X-ray shock emission. Assuming the particle acceleration in the intrabinary shock in spiders is not substantially less efficient than known pulsar wind nebulae (i.e., particles may radiate at up to the synchrotron radiation reaction limit of $m_e c^2 / \alpha_f \sim 70$ MeV where $\alpha_f \approx 1/137$, and $m_e c^2$ is the electron rest energy), the spectral energy distributions of these these systems is predicted to be double-humped, with the synchrotron component peaking  at several hundred MeV and the (external) inverse Compton component at several TeV. Two-photon absorption of TeV photons by the companion photons is found to be generally negligible. 

To match the nonthermal X-ray fluxes, shock magnetic fields of order $\sim 1-10$ G and pulsar pair multiplicities of $\sim 500-1000$ are required. This implies the magnetization parameter $\sigma$ (the ratio of magnetic to plasma energy densities) is of order $10^{-5}-10^{-2}$, commensurate with those required to fit known pulsar wind nebulae. The pair multiplicities found are generally higher than predicted by polar cap pair cascades for millisecond pulsars, and may imply the existence of surface multipolar components \cite[as also recently suggested by {\it{NICER}} results of individual MSPs][]{2019ApJ...887L..23B,2021ApJ...907...63K}.

\begin{figure}[htp]
\centering
\includegraphics[width=0.8\textwidth]{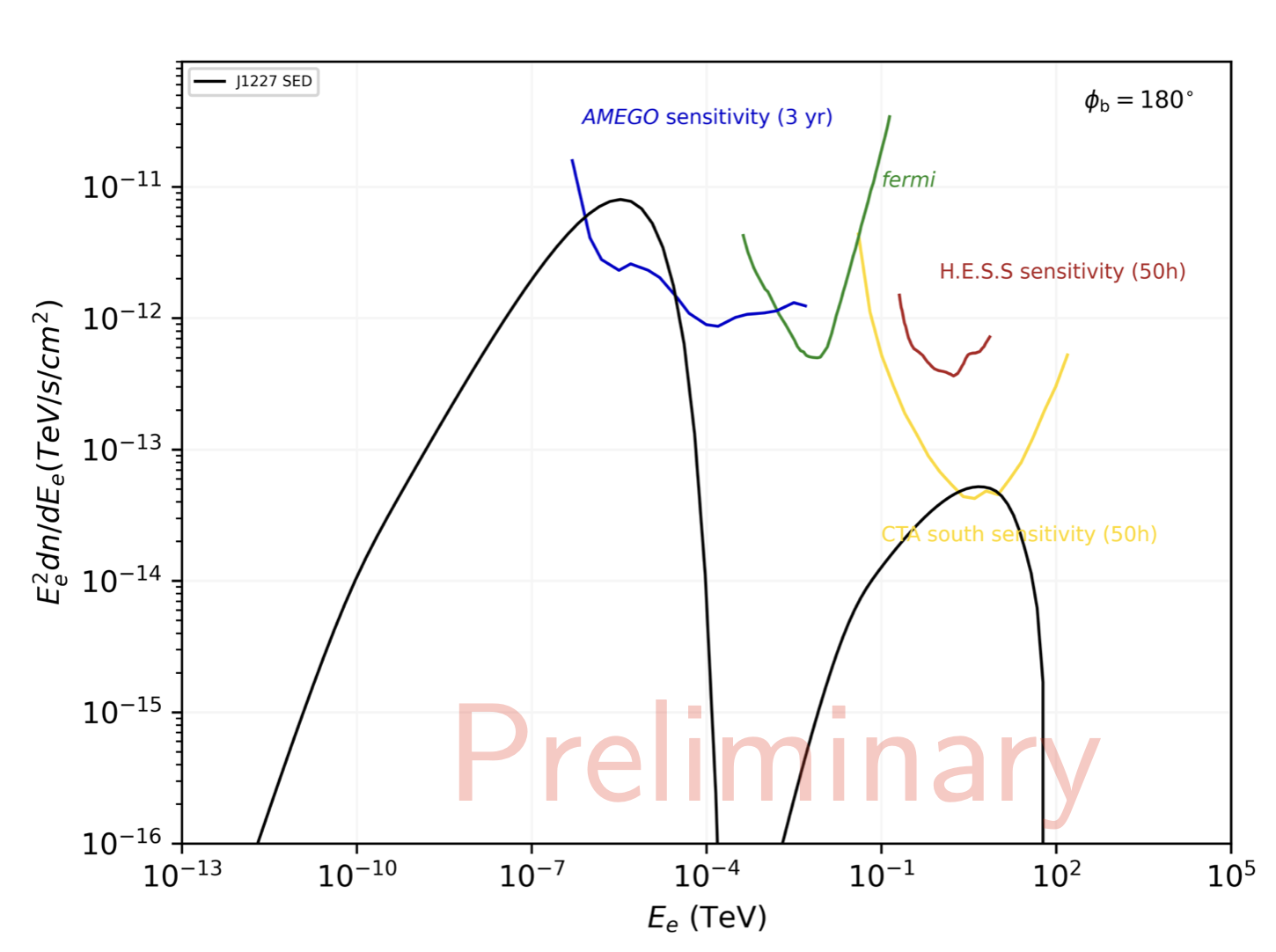}
\caption{Model spectral energy distribution for redback XSS J12270--4859, computed using the {\tt{umbrela}} code, demonstrating that this system is a plausible target for CTA (in addition to those systems, detailed in \cite{2020ApJ...904...91V}).}
\label{fig1}
\end{figure}

An example of such a two-humped spectral energy distribution is shown for  transitional redback PSR J1227--4853 in Figure~\ref{fig1} (reported here for the first time).  PSR J1227--4853 (also known as XSS J12270--4859) is an energetic MSP with an orbital period of 6.91 hr, orbital inclination $\sim 60^\circ$, irradiated companion with mass $\sim 0.3 M_\odot$ and is relatively nearby at distance $\sim 1.4-1.6$ kpc \cite{2014MNRAS.441.1825B,2014ApJ...789...40B,2015ApJ...800L..12R,2015MNRAS.454.2190D,2020MNRAS.492.5607D}. The system exhibits strong nonthermal orbital modulated emission in the {\it{XMM-Newton}} and {\it{NuSTAR}} bands \cite{2015MNRAS.454.2190D,2020MNRAS.492.5607D} with photon index $\Gamma_X \approx 1.2$ at a few$\times 10^{-13}$ erg cm$^{-2}$ s$^{-1}$ flux in the $3-79$ keV band. In Figure~\ref{fig1}, parameters are chosen to roughly match the observed X-ray flux while allowing for detectible TeV emission. Among the parameters, the magnetic field at the shock is set at $0.4$ G, and pair multiplicity at $700$ while the well-constrained binary parameters and inclination are obtained from \cite{2015ApJ...800L..12R,2015MNRAS.454.2190D}.

The predicted spectra in Figure~\ref{fig1} and systems detailed in \cite{2020ApJ...904...91V} are currently somewhat unconstrained solely by available X-ray and optical data -- more broadband detections are needed, especially by IACTs and MeV instruments (e.g., e-ASTROGAM \cite{2018JHEAp..19....1D}, AMEGO \cite{2019BAAS...51g.245M} or GECCO\footnote{Also see Alexander Moiseev's ICRC 2021 proceedings} \cite{2021cosp...43E1372M}), in order to better constrain the model parameter space.  Nevertheless, for nearby systems such as XSS J12270--4859 or those considered in \cite{2020ApJ...904...91V}, the expected MeV flux is readily characterizable by next-generation MeV concepts such as AMEGO or GECCO. For J1723--2837 or J1311--3430, the TeV flux may reach H.E.S.S.-II sensitivity for $\sim 60$ hours (and would be likely detectable by CTA in only a few hours). {\it{Fermi}}-LAT detections of orbitally-modulated multi-MeV emission, whose phasing is consistent within that for the shock, also constrain the models (such as in black widow J1311--3430). Moreover, we find that J1311--3430 may be observable by H.E.S.S.-II if the companion is in a high optical flaring state owing to the target photon number density scaling as $n_\gamma \propto T^3$. IACTs are particularly suited for such short-timescale variability and orbitally phase-resolved observations of these binaries are highly encouraged by current IACTs and CTA.

\bibliographystyle{JHEP}
\bibliography{refs_list}

\end{document}